\documentclass[]{fairmeta}

\usepackage[english]{babel}

\usepackage{amsmath}
\usepackage{amsfonts}
\usepackage{graphicx}

\title{Temporal structure of the language hierarchy within small cortical patches}

\author[1,2*]{Julien Gadonneix}
\author[2,3*]{Mingfang (Lucy) Zhang}
\author[4]{Jérémy Rapin}
\author[1]{Linnea Evanson}
\author[1,2\dagger]{Pierre Bourdillon}
\author[4\dagger]{Jean-R\'emi King}

\affiliation[1]{Hospital Foundation Adolphe de Rothschild}
\affiliation[2]{Paris Cité University}
\affiliation[3]{École Normale Supérieure, Université PSL, CNRS}
\affiliation[4]{Meta AI}

\contribution[*]{Equal contribution}
\contribution[\dagger]{Joint supervision}

\abstract{
Speech production requires the rapid coordination of a complex hierarchy of linguistic units, transforming a semantic representation into a precise sequence of articulatory movements.
To unravel the neural mechanisms underlying this feat, we leverage recordings from eight $3.2 \times 3.2$ mm $64$-microelectrode arrays implanted in the motor cortex and inferior frontal gyrus of two patients tasked to produce twenty thousand sentences.
We show that a hierarchy of linguistic features are robustly encoded in most of these small cortical patches. Contrary to our expectations, instead of a clear macroscopic organization between patches, we observe a multiplexing of phonetic, syllabic and lexical representations within each cortical patch. Critically, this coding scheme dynamically changes over time to allow successive phonemes, syllables and words to be simultaneously represented without interference.
Overall, these results, reminiscent of position encoding in transformers, show how small cortical patches organize the unfolding of the speech hierarchy during language production.
}

\date{\today}
\correspondence{
\email{jgadonneix@u-paris.fr},
\email{lucy.zhang@psl.eu},
\email{pierre.bourdillon@neurochirurgie.fr},
\email{jeanremi@meta.com}}

\begin{document}
\maketitle


\begin{figure}[ht]
    \centering
    \includegraphics[width=\linewidth]{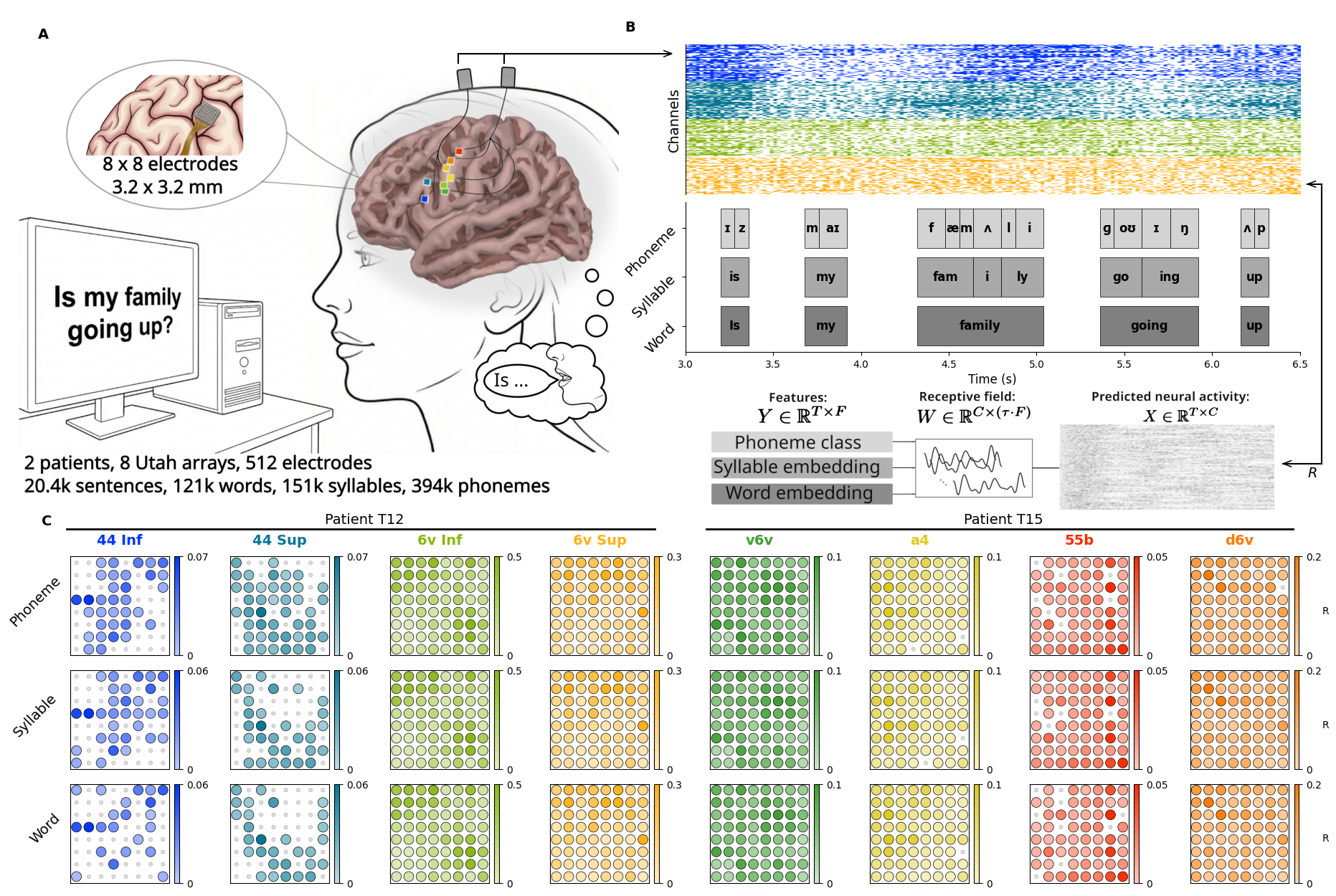}
\end{figure}
\addtocounter{figure}{0}
\begin{figure}[t!]
    \caption{\textbf{Experimental design and microscopic mapping of the speech hierarchy.} \textbf{A}. Intracortical neural activity was recorded from two participants (T12 and T15) implanted with a total of eight $64$-channel microelectrode arrays in the motor cortex and IFG. Participants performed a sentence production task following a visual cue (see \Cref{Method}).
    \textbf{B.} Neural activity and feature hierarchy with an example sentence from T12. Top: Neural activity plot of the binned threshold crossing neural feature for one of the patients (the data is binarized for visualization). Middle: Three levels of representations are analyzed from binned neural activity ($X$): phonetic features ($39$ categories via One-Hot-Encoding), syllabic features (sub-word embeddings via FastText \citep{bojanowski2017enriching}) and lexical features (word embeddings via Spacy \citep{honnibal2020spacy}). Bottom: Performance is summarized using Pearson correlation ($R$) between predicted and actual vectors, where predictions come from the TRF encoding analyses.
    \textbf{C.} Neural activity encoding scores per electrode for phonemes (top), syllables (middle), and words (bottom). Electrodes highlighted in color indicate significant encoding scores ($p \leq 0.05$, one-sided $t$-test across folds), with colors corresponding to the arrays shown in \textbf{A}. Sparse representations of phonemes, syllables, and words are observed in language areas (44, 55b), while dense, highly predictable representations are found in premotor areas (6v, d6v).}
    \label{fig1}
\end{figure}

\section{Introduction}

Speech production requires extraordinarily precise motor control. During this process, the brain translates the meaning of a concept into a rapid, hierarchically-organized sequence of words, while iteratively constructing syllabic and then phonetic sequences that precisely coordinate the respiratory, laryngeal, and articulatory systems.

The brain-level network involved in speech production is increasingly well characterized and is traditionally viewed as a sequence of distinct spatial and functional stages. It transitions from conceptual preparation and word retrieval in the left middle and inferior temporal gyri to motor execution in the primary motor cortex and the supplementary motor area, mediated by phonological information in the posterior superior temporal gyrus \citep{indefrey2011spatial, price2012review, morgan2025decoding}. Along this path, the language network maintains a dedicated set of linguistic representations that are distinct from both lower-level motor mechanisms and higher-level systems of reasoning \citep{fedorenko2024language}. This organization allows the brain to map abstract meanings onto a hierarchy of linguistic features from contextual word representations down to individual phonemes \citep{zhang2025thought}.
However, the precise neural population code coordinating these movements remains poorly understood. This stems from a critical lack of data at the population level, where the fundamental units of speech may be represented.

In this study, we leverage a large intracortical dataset to map the microscopic mechanisms of speech production. We analyze neural activity from two amyotrophic lateral sclerosis (ALS) patients implanted with a total of eight Utah arrays in the motor cortex and inferior frontal gyrus (IFG) \citep{willett2023high, card2024accurate}. This dataset comprises a massive corpus of $20.4$ thousand sentences, $121$ thousand words, $151$ thousand syllables, and $394$ thousand phonemes, providing the statistical power necessary to track subtle neural trajectories.

With this, we consider three main lines of questioning.
First, we evaluate whether the patterns of neural activity reliably represent a variety of linguistic features within cortical patches, using linear encoding analyses. We then employ linear decoding analyses to observe how these representations are distributed across these patches.
Second, we evaluate whether these linguistic features follow a macroscopic gradient across cortical patches, from high-level representations in the IFG down to low-level phonetic features in the motor cortex.
Finally, we evaluate how these neuronal populations rapidly chain successive linguistic features without interference.

\section{Results}

\subsection{Speech representations in local neural populations}

\paragraph{Language representations.}
To investigate how the human cortex coordinates the complex sequence of speech, we first establish a high-resolution mapping of linguistic features onto the neural activity within each grid. Encoding analyses with temporal response functions (TRF) reveal that localized neural populations represent speech features. These neurons exhibit remarkably high predictability for phoneme, syllables, and words, suggesting that the fundamental units of the language hierarchy are already represented at the level of single-unit or multi-unit activity (\Cref{fig1}C).

\paragraph{Within grid overlap in feature encoding.} \label{overlap}
We first investigate the spatial organization of phonetic, syllabic and lexical representations at the microscopic scale to determine if these hierarchical levels are encoded by distinct or shared groups of electrodes within each microelectrode array. 
To test this, we compare the spatial distribution of electrodes that significantly encode low-level phonetic features with those encoding higher-level syllable and word embeddings across all eight Utah arrays. This comparison allows us to test whether the microscopic circuits in the motor cortex and IFG maintain the functional specialization typically observed in whole-brain neuroimaging and electro-corticography (ECoG) studies \citep{flinker2015redefining, silva2022neurosurgical, goldstein2024information, goldstein2025unified}.
Our analysis reveals a striking lack of anatomical segregation between these three levels of processing at the neural population level.
This massive spatial overlap is highly consistent across the microelectrode arrays. To quantify this, we identify the top $10$ best encoded electrodes for phonemes, syllables and words within each individual array and count the number of electrodes simultaneously encoding all three features. Across the Utah arrays, on average the same $9.3$ ($\pm 0.2$ standard error of the mean (SEM) across electrodes) electrodes, which we call the "overlap" within each grid are significantly involved in representing these three levels of the speech hierarchy (one-sided Wilcoxon signed-rank test on the overlap across microelectrode arrays, $p < 0.005$).
These results suggest that these microscopic cortical patches function as a neural "mosaic" where information from different speech units are simultaneously represented in the same neurons (more details about this model in \Cref{mosaic}). By multiplexing several layers of the language hierarchy within the same physical space of localized neural populations, the brain achieves a high degree of information optimization, maximizing the representational capacity of a limited population of neurons.

\paragraph{Neuronal tuning.}
Many neurons have specialized roles in visual or auditory processing \citep{chan2014speech, leonard2024large, gerken2024decoding}.
Here, we identify neurons that exhibit precise tuning for particular phonemes, syllables or words (\Cref{fig2}C-E, see \Cref{Method}), which was expressed by a higher spiking activity than for other events.  
This aligns with the idea of specialized roles and offers early evidence for the feasibility of decoding complex sentences from a small number of neurons.

\subsection{Anatomical and temporal feature organization across grids}
To determine the precise anatomical and temporal organization of speech features during an attempted speech task, we shift our analysis from the localized populations to a comparative evaluation of decoding performance between individual Utah arrays across different cortical regions and linguistic levels.
This analysis allows us to replicate the speech hierarchical structures identified in previous literature \citep{evanson2025emergence, goldstein2025unified, zhang2025thought} while extending these findings to the resolution of individual neurons, revealing how high-resolution features are coordinated within localized cortical patches.

\begin{figure}[t!]
    \centering
    \includegraphics[width=\linewidth]{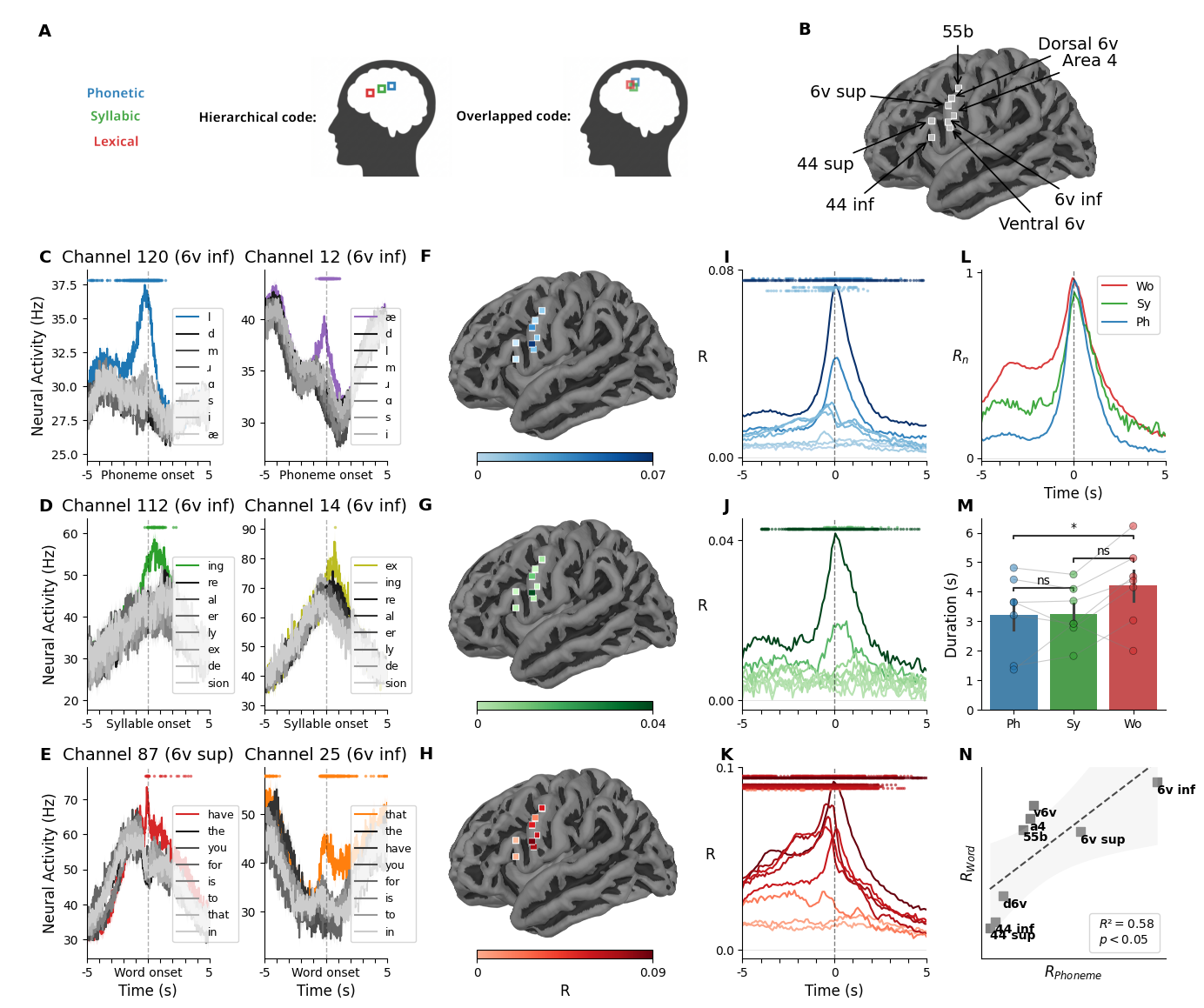}
    \caption{\textbf{High-resolution encoding and decoding of the speech hierarchy in the motor cortex and IFG.} \textbf{A.} Competing anatomical hypotheses for the speech hierarchy. Left: A hierarchical segregation model, where distinct cortical regions (blue for phonetic, green for syllabic and red for lexical) code for different linguistic levels. Right: The overlapping neural mosaic model, where all representations are encoded in the same localized neural populations.
    \textbf{B.} Cortical map of all the Utah arrays.
    \textbf{C-E.} Neural activity in Hz from representative electrodes aligned to the onset of phonemes (\textbf{C}), syllables (\textbf{D}) or words (\textbf{E}), demonstrating sharp tuning to linguistic units at different levels; significance is determined by two-sided Mann-Whitney $U$-tests across splits for $p < 10^{-4}$.
    \textbf{F-H.} Cortical map of phoneme (\textbf{F}), syllable (\textbf{G}) or word (\textbf{H}) peak decoding score within a $10$ s temporal window around onset, showing the eight Utah arrays. 
    \textbf{I-K.} Time-resolved decoding scores ($R$) for phonemes (\textbf{H}), syllables (\textbf{I}) or words (\textbf{J}) for all eight Utah arrays; significance is indicated by dots at each timestamp and array, based on one-sided $t$-tests across splits for $p < 10^{-11}$ ($10^{-7}$ for syllables). Color coding is consistent with panels \textbf{F-H}.
    \textbf{L.} Normalized decoding scores from the 6v inf array only for words, syllables and phonemes.
    \textbf{M.} Comparison of representation duration, showing that word-level information persists significantly longer than phoneme-level information ($t_{Wo} - t_{Ph} = 0.99 \pm 0.15$ s; $p < 0.05$, two-sided Wilcoxon signed-rank test and SEM across electrodes with FDR correction). Each triplet of dots represents a Utah array. Ph: phoneme. Sy: syllable. Wo: word.
    \textbf{N.} Linear regression of word versus phoneme peak decoding performance across arrays ($p < 0.05$, two-sided Wald test with $t$-distribution; $R^2 = 0.58$).
}
    \label{fig2}
\end{figure}

\paragraph{The role of Broca's area.}
Our results highlight a known dissociation between the motor cortex (arrays 6v inf, 6v sup, v6v, a4, 55b and d6v, \Cref{fig2}B) and Broca’s area in the IFG (arrays 44 inf and 44 sup) \citep{flinker2015redefining, silva2022neurosurgical, willett2023high}. Encoding scores for phonemes, syllables and words were significantly greater in the motor cortex than in Broca’s area, which did not exhibit a similarly robust involvement for these specific linguistic events ($p < 0.0005$, two-sided Mann-Whitney $U$-test comparing the number of significantly encoded electrodes in the two areas, \Cref{fig1}C). This suggests that while Broca’s area is traditionally associated with high-level language production, the actual implementation of the rapid, sequential dynamic code for phonemes, syllables and words may be more concentrated in the circuits of the motor cortex.
This discrepancy raises a fundamental question about the spatial organization of the speech hierarchy: how, then, are these linguistic representations distributed across the broader cortical landscape?

In the results presented from \Cref{fig2} onward, the analyses of syllables were restricted to those occurring in plurisyllabic words. This choice was made to differentiate syllabic patterns from those of monosyllabic words, which are prevalent in the dataset. Analyses for all syllables, including those in monosyllabic words, are available in the supplementary information (\Cref{supp2}-\Cref{supp4}).

\paragraph{Anatomical specialization.} \label{mosaic}
To investigate the spatial organization of the speech hierarchy, we compare two anatomical hypotheses: hierarchical segregation and the overlapping neural "mosaic" (\Cref{fig2}A). The segregation model posits that distinct cortical regions are specialized for different linguistic levels, typically placing phonetic processing in motor areas and lexical-semantic processing in the IFG. In contrast, the overlapping model suggests that all levels are represented within the same localized neural populations.
As the syllabic analysis was limited to syllables within plurisyllabic words, the sample size was reduced from $151,000$ to $55,000$ observations. This reduction introduced greater variability in the decoding analyses across most arrays. Consequently, syllabic interpretation was excluded from certain subsequent analyses and conclusions.
\\
A critical observation is the performance variance across regions (\ref{fig2}F-H). While all Utah arrays in the motor cortex exhibit a significant peak in decoding scores of phonetic and lexical information, the decoding scores of the arrays located in the IFG showed a lack of peak performance across time (\Cref{fig2}I-K). To determine the significance of these peaks, we compute the relative increase in decoding performance between the "half-peak" (see \Cref{Method}) and $5$ seconds after the event onset.
In the superior portion of area 44 (44 sup), phonetic information does not induce a significant peak in decoding scores ($\Delta = 23 \pm 12 \%$; $p > 0.05$, one-sided paired-sample $t$-test and SEM across splits). While word-level information is statistically significant in this region, the relative increase is notably low ($\Delta = 29 \pm 8 \%$; $p < 10^{-6}$) compared to the motor cortex ($\Delta > 125\%$). In the inferior portion of area 44 (44 inf), we observe a similar trend where linguistic representations are significant but characterized by a low relative difference ($\Delta = 50 \pm 12 \%$ and $\Delta = 66 \pm 12 \%$; $p < 5 \times 10^{-4}$).
These results suggest that while the motor cortex is a primary hub for both articulatory and lexical features, the IFG's contribution to speech production may involve higher-order structural processes or different temporal scales that are not fully captured by onset-locked linear decoding.

\paragraph{Distinct temporal dynamics across the speech hierarchy.}
Beyond this spatial distribution, the linguistic hierarchy is also distinguished by the temporal scale of its neural representations.
We analyze the temporal dynamics of the different speech units by calculating the half-peak to half-peak duration of decoding accuracy.
While these hierarchical differences in temporal persistence are marginal, they provide a critical initial indication of the distinct time scales at play within the cortical patches. We find that lexical features are longer-lasting than phonetic ones across the majority of recording sites (\Cref{fig2}M). Word-level representations persist significantly longer than phoneme-level ones ($p < 0.05$; $t_{Wo} - t_{Ph} = 0.99 \pm 0.15$ s, two-sided Wilcoxon signed-rank test and SEM across electrodes with false discovery rate (FDR) correction), reflecting the longer natural duration of lexical units in the speech stream \citep{goldstein2025unified, gwilliams2025hierarchical, evanson2025emergence}. For this comparison, the Utah array in the superior part of area 44 was excluded from the analysis, as phonetic information does not peak significantly in this region (\Cref{fig2}F, I).
The increase in neural representation duration appears to reflect the hierarchical organization of speech. Specifically, when examining the microelectrode array in the inferior region of area 6v (\Cref{fig2}L), a progressive lengthening of peak decoding duration is observed, transitioning from words to phonemes via syllables.
Although subtle, this trend reflects the longer natural duration of lexical units in speech sequences.
\\
Furthermore, word decoding in the motor cortex exhibits a distinct early peak before word onset, giving the decoding curve a bimodal appearance (\Cref{fig2}K). This early component coincides temporally with the onset of the visual cue (\Cref{supp1}), suggesting it reflects the initial perception of the sentence and the lexical retrieval of the prompt. Notably, this perceptual peak is absent in the phonetic decoding curve, which only rises during the motor preparation and execution of speech.

\paragraph{The shared neural substrate.}
To test whether phonetic and lexical features are simultaneously represented within the same microscopic circuits, we evaluate the relationship between their decoding performance at the resolution of individual Utah arrays.
We find that the Utah arrays that are most effective at decoding phonemes are the same arrays that best decode word embeddings ($p < 0.05$, two-sided Wald test with $t$-distribution; $R^2 = 0.58$; \Cref{fig2}N).
The high correlation between these decoding performances confirms that the same microscopic cortical patches represent multiple levels of the language hierarchy. This result directly improves our earlier encoding analysis (\Cref{overlap}), which demonstrated a massive spatial overlap of these features within individual microelectrode arrays. Rather than showing anatomical segregation, our findings demonstrate that a small population of neurons in the motor cortex can simultaneously embed a rich representation of both articulatory building blocks and lexical meaning during speech production.

These findings compare two seemingly contradictory organizational principles of the speech hierarchy at the microscopic scale. While we observe a clear regional dissociation with the motor cortex acting as the primary hub for all speech features compared to the relatively sparse involvement of the IFG, the organization of these features across the microelectrode arrays defines a shared neural substrate rather than anatomical segregation. Furthermore, a consistent temporal trend emerges where higher-level lexical units are maintained significantly longer and established earlier than their phonetic counterparts. Together, these results demonstrate that the motor cortex provides a unified, high-dimensional space where multi-scale linguistic features are simultaneously represented to coordinate the fluid unfolding of speech.

\subsection{Temporal sustainment of speech representations}

\begin{figure}[t!]
    \centering
    \includegraphics[width=\linewidth]{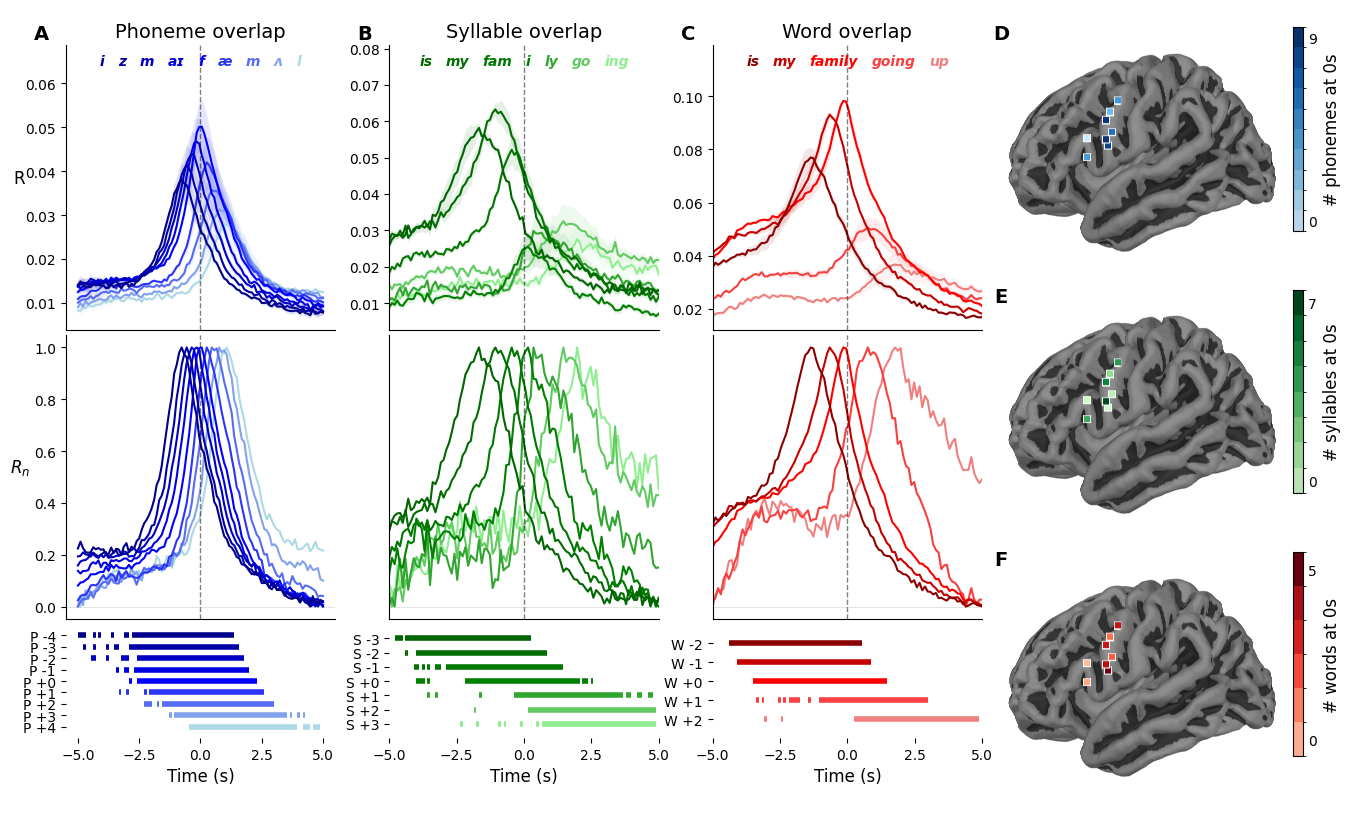}
    \caption{\textbf{Hierarchical overlapping neural representations.} \textbf{A-C.} Overlap of decoding scores in neural representations for successive phonemes (\textbf{A}), syllables (\textbf{B}) or words (\textbf{C}) aligned to the onset of the current event. These plots represent scores (top) and normalized scores (middle) for decoders trained on the $i^{th}$ ($i \in \{-4, -3, ..., 4\}$) phoneme, syllable or word but time-locked on the $0^{th}$. Colored bars (bottom) represent the values above the median. Results are derived from all electrode data and averaged across patients, with error representing the SEM. The example phonemes, syllables and words at the top are used for illustration purposes.
    \textbf{D-F.} Cortical maps showing the number of phonemes (\textbf{D}), syllables (\textbf{E}) and words (\textbf{F}) significantly decoded simultaneously at $t = 0$ s for each Utah array.
}
    \label{fig3}
\end{figure}

\paragraph{Simultaneous representations of past, present and future.}
We have established that phonetic, syllabic and lexical features occupy the same physical space within and across Utah arrays over time scales of seconds. However, a fundamental challenge in neural sequencing is maintaining the identity of successive items without interference. The average phoneme, syllable and word durations are $326 \pm 55$ ms, $518 \pm 148$ ms and $618 \pm 296$ ms respectively, whereas we observe that the average persistence of their representations in the neural space is much longer (\Cref{fig2}K). This raises the following question: do representations of successive speech features temporally overlap?
\\
We therefore test whether the microscopic populations of each electrode array represent multiple units of the speech stream simultaneously.
By training ridge regression models on neural signals to predict the current, previous and subsequent events of the speech hierarchy, we observe a significant temporal overlap in their neural representations (\Cref{fig3}A-C). This phenomenon already observed at the whole-brain scale \citep{gwilliams2022neural, gwilliams2025hierarchical, zhang2025thought} is also found here in a small neural population.
Our results show that the neural representation of a phoneme, a syllable or a word starts long before and persists long after its motor execution begins. This occurs while the representation of the previous unit is still encoded and the upcoming unit is already active in the same population. Rather than a discrete handover from one unit to the next, the neural activity at any given moment during speech production contains information about several consecutive linguistic events.

\paragraph{Overlapping at the microscopic scale.}
Crucially, this simultaneous representation is not merely a global property of the motor cortex but is present at the microscopic scale of individual Utah arrays (\Cref{fig3}D-F). On these tiny cortical patches, we find that the same few neurons represent the current phoneme and simultaneously the preceding and succeeding units. This indicates that local circuits do not process speech as isolated events. Instead, they implement a form of temporal multiplexing, where the past, current, and future states of the speech hierarchy are nested within the same neural mosaic. 
This finding suggests that part of the computations required to manage sequences of speech events is contained within these small cortical patches.

In our analysis of syllable and word sequences, we observe an asymmetry between the decoding of previous and next units (\Cref{fig3}C). This is particularly evident in the second patient’s data, where a dataset imbalance ($> 18$\% of sentences beginning with the word "I") leads to a bias in the predictive window for preceding lexical items (as those items are likely to be the first word of the sentence, see \Cref{supp5}). Despite this artifact of the corpus, the broader trend remains consistent: the microscopic neural code maintains a multi-unit buffer that embeds successive elements in the speech hierarchy and may coordinate the transitions.

\subsection{A hierarchy of dynamic neural codes}

\begin{figure}[t!]
    \centering
    \includegraphics[width=\linewidth]{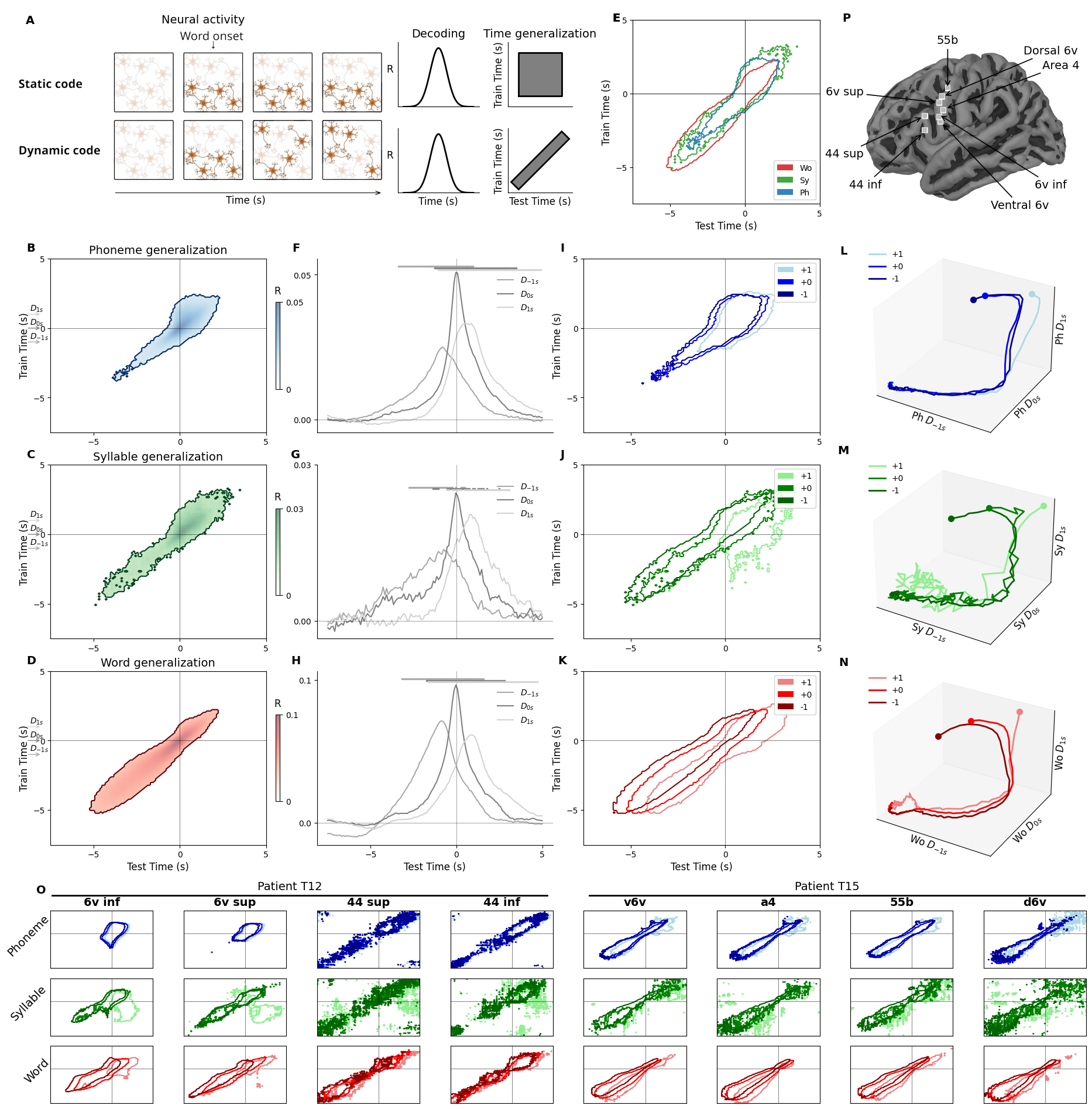}
    \caption{\textbf{Dynamic neural trajectories coordinate the speech hierarchy.} \textbf{A.} Schematic representation of static versus dynamic neural codes. In a static code, a feature is represented by a fixed pattern of activity, predicting a square time-generalization matrix; in a dynamic code, the representation evolves over time, predicting a diagonal matrix. 
    \textbf{B-D.} Temporal generalization matrices for phonemes (\textbf{B}), syllables (\textbf{C}) and words (\textbf{D}). Contours enclose scores exceeding the mean by $1.5$ standard deviations.
    \textbf{E.} Same as \textbf{B–D}, combined in a single plot.
    \textbf{F-H.} Performance of decoders trained at specific time points ($D_{-1s}$, $D_{0s}$, $D_{1s}$) and tested across time for phonemes (\textbf{F}), syllables (\textbf{G}) and words (\textbf{H}); significance is determined by two-sided Wilcoxon signed-rank tests across splits for $p < 10^{-10}$.
    \textbf{I-K.} Temporal generalization matrices for phonemes (\textbf{I}), syllables (\textbf{J}) and words (\textbf{K}) for successive speech units ($+1$, $0$, $-1$). Contours enclose scores exceeding the mean by $1.5$ standard deviations.
    \textbf{L-N.} $3D$ visualization of the neural code trajectories until $1.5$ s after phoneme (\textbf{L}), syllable (\textbf{M}) or word (\textbf{N}) onset for successive speech units ($+1$, $0$, $-1$). The axes represent normalized decoding scores for decoders trained at specific time points ($D_{-1s}$, $D_{0s}$, $D_{1s}$).
    \textbf{O.} Temporal generalization for each individual Utah array. Contours enclose scores exceeding the mean by $1.5$ standard deviations.
    \textbf{P.} Cortical map of all the Utah arrays (reproduced from \Cref{fig2}B for clarity).
    }
    \label{fig4}
\end{figure}

\paragraph{The evolving neural code.}
We observed that the neural representations of successive linguistic units exhibit a temporal overlap which raises a fundamental computational problem: how does the brain coordinate these co-active signals without mutual interference?
By employing temporal generalization analysis \citep{king2014characterizing}, we inspect the trajectory through the neural state space of single speech units by training linear decoders at each time step relative to the onset ($t_{train}$) and tested their performance across all other time steps ($t_{test}$) between $7.5$ seconds before and $5$ seconds after event onset (see \Cref{Method}).
There are two competing hypotheses for how linguistic information is maintained in neural populations during speech production (\Cref{fig4}A). In the static coding model, a speech unit is represented by a fixed, stable pattern of neural activity that persists throughout the event's duration \citep{gwilliams2025hierarchical}. This would result in a square-shaped temporal generalization matrix, as a decoder trained at any single time point would successfully generalize at all other time points.
In contrast, the dynamic coding model proposes that the neural representation of a phoneme, syllable or word evolves along a specific trajectory \citep{king2014characterizing, gwilliams2022neural, gwilliams2025hierarchical, zhang2025thought}. This dynamic evolution would result in a diagonal temporal generalization matrix, indicating that a neural pattern at one moment is transient and quickly transitions into a different, non-interfering state.
The resulting generalization matrices (\Cref{fig4}B-D) exhibit a clear diagonal profile. While the identity of a phoneme is present in the neural population for a relatively long duration as seen in the results from \Cref{fig2} and \Cref{fig3}, the specific neural pattern used to represent that phoneme at one moment in time is only valid for a relatively short window of time ($1.03 \pm 0.74$ s, mean and standard deviation computed for $-3$ s $< t_{train} < 1$ s), and similarly for syllables and words ($1.64 \pm 0.35$ s and $2.45 \pm 0.27$ s, mean and standard deviation computed for $-3$ s $< t_{train} < 1$ s) with a gradual increase. A decoder trained at a fixed time step of $1$ second before onset fails to generalize to the activity $1$ second after onset, even though the same speech unit is still being processed (\Cref{fig4}F-H).
This indicates that over the duration during which a speech unit is represented, the neural code evolves. This rapid evolution enables the cortex to distinguish between the same linguistic feature at different stages of its production, like during planning, execution and memory, effectively keeping track of the relative order of items in a sequence.

\paragraph{Interference reduction via different subspaces.}
The functional advantage of this dynamic code lies in its ability to represent sequences of the speech hierarchy by evolving the neural signature of a phoneme over time.
Specifically, we find that the representation of a past phoneme, which has moved to a later stage of its trajectory, occupies a different neural subspace from that of the current or future phonemes because the trajectories are parallel but offset (\Cref{fig4}I-K). Indeed, all phonemes, syllables and words follow a similar trajectory in terms of neural code. They reuse the same dynamical patterns to move through the sequence (\Cref{fig4}L-N).
This systematic movement allows several speech units of the same hierarchical level to be represented simultaneously with the same evolving neural code during their mental representations, but with a time lag, illustrating how the cortex reuses a consistent dynamical pattern for each unit in the sequence. This creates an ordered sequence of language blocks without their signals interfering with one another. By transforming time into a spatial dimension within the neural population activity, the brain can represent multiple units simultaneously without them overlapping in the same representational space. The multiplexing of signals from consecutive speech events is achieved through the efficient reuse of the same evolving localized neural populations to embed speech units.
\\
In contrast, under the static code hypothesis, speech units are treated as isolated events; however, because of the overlap in time of successive events (\Cref{fig3}), such a static representation would lead to signal interferences of the sequences within the brain.

\paragraph{Comparison of speech hierarchical levels.}
We observe distinct dynamic patterns for phonemes, syllables and words, with their temporal characteristics diverging in two key ways. First, in line with the temporal scales observed previously, the maintenance windows exhibit a hierarchical gradient: the duration increases from phonemes to words (\Cref{fig4}E).
This stepwise increase in temporal stability suggests that the local neural representations are maintained for periods proportional to the complexity of the linguistic unit, mirroring the nesting of features within the speech hierarchy.
Second, our analysis of the neural state-space trajectories reveals a significant difference in the velocity of neural representation dynamics. We quantified the evolution speed as the angle of the decoding diagonal (see \Cref{Method}), where a processing trajectory that moves faster than the training data results in an angle of less than $45^\circ$ compared to the $y$-axis. Slower trajectories produce an angle greater than $45^\circ$. We find that this velocity follows a clear hierarchical gradient, with the speed for words ($44.8 \pm 0.2 ^\circ$) being significantly lower than for syllables ($42.5 \pm 0.5 ^\circ$), which is in turn lower than for phonemes ($35.4 \pm 2.1 ^\circ$) ($p < 0.005$, two-sided Wilcoxon signed-rank tests and SEM across splits and subjects with FDR correction).
This suggests that while all levels utilize the same trajectorial mechanism, the velocity of the dynamic code is tuned to the specific linguistic hierarchical level of the unit, with phonemes requiring more rapid state transitions to accommodate the high-speed demands of speech motor control.
\\
Additionally, we note a clear asymmetry in the temporal scales: neural representations are present significantly longer before the onset of the event than after. Once again, there is a hierarchical shift for this anticipatory processing, where the difference between pre-onset and post-onset decodability duration increases stepwise from phonemes to words. Specifically, this pre-onset bias was $2.21$ s for phonemes ($SEM = 0.16$), $2.34$ s for syllables ($SEM = 0.09$) and $3.07$ s for words ($SEM = 0.15$), with each level exhibiting a significant lead time ($p < 10^{-4}$ for each level individually, two-sided Wilcoxon signed-rank tests across splits and subjects).
This gradient confirms that higher-level linguistic structures are established in the local circuit before their constituent syllabic and phonetic components. It suggests that the brain must engage in extensive lexical planning and retrieval before motor execution begins.

\paragraph{Microscopic implementation of a global mechanism.}
Crucially, we observe this dynamic code within each individual Utah array, with all eight micro-electrode arrays across both patients exhibiting the characteristic diagonal temporal generalization profile (\Cref{fig4}O). This pattern remains robust even in the IFG. Although these regions yield overall smaller decoding scores, the underlying temporal dynamics are identical to those observed in the motor cortex.
The fact that this sophisticated dynamic code, typically associated with macroscopic network activity \citep{gwilliams2022neural, gwilliams2025hierarchical, zhang2025thought}, is fully implemented within these localized microscopic patches suggests that even small circuits of the human cortex implement the complex mechanism required to coordinate the rapid, hierarchical flow of language.
Despite the lower decoding scores in the IFG (\Cref{fig1}-\Cref{fig3}), temporal generalization analysis reveal that this cortical area still implements a robust dynamic neural code. The characteristic diagonal pattern in the temporal generalization matrices (\Cref{fig4}O) is present across both area 44 sites, mirroring the dynamics found in the motor cortex. This indicates that while the IFG may not represent individual speech units with the same fidelity as the motor cortex during execution, it utilizes the same underlying hierarchy of dynamic codes to evolve neural representations over time.

\section{Discussion}

\paragraph{Results summary.}
These results show with a large number of high-resolution neural recordings that a variety of linguistic features are simultaneously represented in the populations of $3.2 \times 3.2$mm cortical patches during language production.
Critically, we demonstrate that these representations do not exist as isolated states but are instead coordinated through a hierarchy of dynamic neural codes. By transforming the relative position of units into specific neural trajectories, the brain can effectively multiplex phonetic, syllabic and lexical features within the same, localized neural populations while avoiding signal interference.

\paragraph{A hierarchy of linguistic features.}
The decodability of these representations strengthens earlier observations with fMRI \citep{hu2023precision}, ECoG \citep{stephen2023latent, morgan2025decoding, liu2025speech, zhang2026human}, single-cell recordings \citep{chan2014speech, leonard2024large} and magneto-encephalography (MEG) \citep{zhang2025thought}. In particular:
\begin{itemize}
    \item While previous high-performance brain-computer interfaces (BCI) using ECoG \citep{chartier2018encoding, anumanchipalli2019speech, metzger2023high} have achieved record-breaking decoding of speech, these studies focused primarily on articulatory mapping, treating speech production as a series of isolated motor events to decompose speech spectrograms. These works prioritized the "what" of speech by linking cortical activity to physical movement. Our work extends these findings by uncovering the "how", demonstrating that a complex hierarchical and sequential organization exists within microscopic cortical patches.
    \item Our results build upon the sliding representation model of speech perception \citep{gwilliams2022neural, gwilliams2025hierarchical}, which shows how the brain buffers a sequence of incoming sounds and speech features. We demonstrate that speech production utilizes a similar dynamic mechanism, but with a fundamental shift in temporal orientation. Unlike the retrospective nature of perception, production is uniquely prospective, requiring the brain to coordinate high-level lexical goals and their phonetic decomposition long before the motor execution begins. 
    \item We extend recent evidence of hierarchical superposition found in typing tasks \citep{zhang2025thought} to the more rapid and fluid domain of language production. While typing involves a series of discrete motor targets, speech requires the seamless chaining of overlapping articulatory gestures. We show that the same dynamic neural trajectories found in typing are utilized here to represent sequences of the different levels of the speech hierarchy.
\end{itemize}
This study provides the first microscopic evidence that the human speech hierarchy is coordinated through a dynamic neural mosaic within localized cortical patches. We demonstrate that a small population of neurons in the motor cortex multiplexes phonetic, syllabic and lexical features through rapidly evolving neural trajectories. We show that these linguistic levels have distinct temporal durations, with lexical information persisting longer. Finally, we reveal that even the IFG (Broca’s area), despite lower representational fidelity, utilizes this same dynamic process to track the relative positions of speech units, confirming that this trajectorial coordination is a fundamental property of the microscopic neural networks that enable fluid human communication.

\paragraph{Hierarchical overlap.}
Contrasting with traditional modular accounts \citep{indefrey2011spatial, price2012review, flinker2015redefining, morgan2025decoding, silva2022neurosurgical}, we do not observe a clear functional segregation between the IFG, typically associated with high-level compositional processes, and the motor cortex, associated with phonetic execution. 
Instead, local neural populations sensitive to low-level articulatory features are equally robust in encoding higher-level information. While this observed overlap may be influenced by limited spatial coverage, it highlights that a variety of hierarchical features co-exist in the high dimensions of micro-cortical circuit activations.
Furthermore, this lack of anatomical segregation may be a reflection of our specific choice of features; because phonemes, syllables and words are intrinsically nested within the linguistic hierarchy, their neural representations likely share a degree of structural dependency that promotes their co-localization within the same population.
This co-localization is accompanied by a high degree of redundancy across the hierarchy, with nearly all recorded electrodes yielding significant information for all linguistic features. The fact that this hierarchical information is reliably present even when sampling from only a small number of neurons suggests that these representations are not sparsely distributed, but are instead redundantly represented throughout large cortical areas.

\paragraph{Temporal overlap.}
Crucially, this representational overlap extends to the temporal domain. Specifically, successive speech events can be simultaneously read-out from these small patches of cortex. 
This phenomenon can be surprising at first, as a naive coding scheme would associate each feature to a unique dimension of the neuronal activation. However, this would lead to catastrophic interference between successive events. 
Our temporal generalization analyses show that these features are, in fact, represented in a continuously changing subspace of the neuronal activity. 

\paragraph{Related work dynamic coding.}
This dynamic coding scheme is consistent with the recent macroscopic MEG results observed during both speech perception \citep{gwilliams2022neural, gwilliams2025hierarchical} and keyboard typing \citep{zhang2025thought}. The present study further demonstrates that this coding scheme is not an artifact of inter-region dynamics, but in fact already occurs in small cortical patches.
More generally, this dynamic coding scheme is functionally analogous to the positional embeddings in sequence transformers \citep{vaswani2017attention}. In these artificial architectures, the model must process tokens in parallel while maintaining their strict sequential order. It achieves this by tagging each token embedding with a unique positional vector that projects the information into a specific coordinate space. By ensuring that successive tokens occupy nearly orthogonal dimensions, the transformer prevents signal collision and maintains the compositionality of the sequence.
\\
Our findings suggest that the human brain implements a biological version of this computational strategy with relative positions \citep{su2024roformer}. By rotating the neural representation into a rapidly evolving trajectory, the small cortical patch effectively tags each phoneme or word with its relative position in the speech stream. This ensures that multiple units can co-exist within the same neural population without mutual interference, maintaining the compositionality required to chain discrete linguistic elements into a continuous, ordered utterance.

\paragraph{Different temporal dynamics across the hierarchy.}
While word-level information persists significantly longer than the other features (\Cref{fig2}, \Cref{fig4}), the velocity of neural state evolution also varies across hierarchical levels, following a clear gradient where the speed for words is significantly lower than for syllables, which is in turn lower than for phonemes (\Cref{fig4}). Interestingly, this velocity is uniform across the cortex and we observe similar speeds of neural trajectories across different micro-electrode arrays for phonetic, syllabic and lexical information (one-way ANOVA, $p > 0.05$ for phonemes, syllables and words).
This suggests that the "intrinsic timescale" \citep{murray2014hierarchy} of the neural code is not a byproduct of varying regional processing speeds, but rather a reflection of the linguistic scale (the software) being consistently mapped onto different anatomical locations (the hardware). While the underlying processing speeds remain uniform across cortical locations, the local dynamic coding adapts to the specific complexity of the speech unit. This creates a flexible framework where the same hardware can support the rhythmic chaining of articulatory movements, varying the temporal window of integration to match the features required for fluid speech.
\\
The fact that the same microscopic neural populations are shared across hierarchical levels raises the possibility that the prolonged maintenance of lexical representations is simply an artifact of phonetic sequencing. Specifically, different segments of a word embedding could be being decoded as their corresponding phonemes are being produced over time. Under this view, our lexical decoding would merely be the summation of its underlying phonetic parts. However, this interpretation is refuted by our finding that word-level information becomes decodable significantly before phonetic information. This temporal lead indicates that a stable lexical representation is established and maintained in the local circuit prior to its decomposition into phonetic and articulatory motor commands, supporting a truly generative and prospective hierarchical process.

\paragraph{Conclusion.}
Our findings provide a high-resolution window into the population-level dynamics that underpin speech motor control. By resolving the fine-grained representations within local cortical ensembles, this work elucidates how hierarchical linguistic features may be transformed into the coordinated motor commands of the vocal tract.

\section{Methods} \label{Method}

\subsection{Protocol and preprocessing}
\paragraph{Data.}
Data were collected from two participants with amyotrophic lateral sclerosis (ALS) enrolled in the BrainGate2 pilot clinical trial \citep{willett2023high, card2024accurate}. Each participant was implanted with four $64$-electrode ($8 \times 8$) microelectrode arrays (Utah arrays), each measuring $3.2$mm by $3.2$mm with a $1.5$mm depth into the cortex. In the first patient, two arrays were situated in the premotor cortex and two arrays were placed in the posterior IFG, specifically targeting Broca’s area. The second patient had two arrays in the premotor cortex, one in the primary motor cortex and one in the middle precentral gyrus. These arrays recorded spiking activity at a microscopic scale, capturing threshold crossings based on the root mean square and binned spike power, allowing for the analysis of activity from single or small groups of neurons.

\paragraph{Ethics statement.}
The data used in this study were collected as part of the BrainGate2 pilot clinical trial ($NCT00912041$). The research was conducted under an investigational device exemption from the U.S. Food and Drug Administration (IDE $\#G09003$). All procedures were also approved by the institutional review board of Stanford University (protocol $\#20804$). Participants provided written informed consent prior to all study procedures.

\paragraph{Data availability.}
The raw neural recordings, annotations, and behavioral data supporting the findings of this study are publicly available on Dryad. For patient T12, the dataset consists of the full, continuous neural recordings spanning the entire experimental sessions\footnote{\url{https://datadryad.org/dataset/doi:10.5061/dryad.x69p8czpq}}. For patient T15, the available data corresponds to the segmented neural trials specifically curated for the Brain-to-Text '$25$ competition \footnote{\url{https://datadryad.org/dataset/doi:10.5061/dryad.dncjsxm85}}. Detailed metadata regarding both datasets can be accessed through these repositories or in the original reports characterizing these neural populations \citep{willett2023high, card2024accurate}.

\paragraph{Task.}
Participants performed a visually cued speech production task with an instructed delay between the displaying of the sentence on a screen and the signal to start the speaking attempt \citep{willett2023high, card2024accurate}. The patients' condition prevents them from producing intelligible speech and so, in addition to vocalized speech, participants performed "mouthing" trials where they attempted articulatory movements without vocalization, resulting in little to no auditory feedback. Previous analyses of this data indicated that decoding results were similar in both conditions \citep{willett2023high}. Each participant produced approximately $10,200$ sentences, yielding a massive corpus of approximately $121,000$ words, $151,000$ syllables, and $394,000$ phonemes (\Cref{fig1}A).

\paragraph{Preprocessing.}
Neural activity was processed by the dataset authors by binning threshold crossings and spike power into discrete time windows. To detect spiking activity, a voltage threshold set at $-3.5$ times the root-mean-square (RMS) of the raw signal was applied on each channel \citep{willett2023high, card2024accurate}. We utilize single-cell peri-stimulus time histograms-like (PSTH) visualizations to study the response of individual electrodes to specific linguistic events, providing the average activity of neural units time-locked to specific speech units.
To ensure precise temporal alignment between the neural signal and the linguistic hierarchy, we annotate the phoneme labels for each session. We retrain the deep learning models (utilizing a Connectionist Temporal Classification (CTC) loss \citep{graves2013speech}) provided by the dataset authors and evaluate these models on our specific dataset to extract fine-grained phoneme labels and timings. Word- and syllable-level labels are then constructed by integrating these extracted phoneme timings with the ground-truth text of the sentences, ensuring that the hierarchy remains synchronized across all levels of representation. We utilized the Spacy Syllables \citep{honnibal2020spacy} and g2pE \citep{g2pE2019} libraries to identify and extract syllable and phoneme boundaries for each word.

\subsection{Language representations}
To investigate the hierarchy of speech production, we extract three levels of linguistic features and we transform each of the three levels of the language hierarchy into target vector representations (\Cref{fig1}B), using features derived from text strings.

\paragraph{Word representations: Spacy.}
At the lexical level, to capture high-level semantic information, we use word embeddings extracted from the Spacy pretrained model \texttt{en\_core\_web\_lg} \citep{honnibal2020spacy}. These embeddings represent words as continuous vectors in a high-dimensional semantic space ($d = 300$), allowing us to test for higher-level semantic representations.

\paragraph{Syllable representations: FastText.}
At the syllabic level, to capture the underlying computational properties, we featurize the extracted syllables using FastText sub-word embeddings \citep{bojanowski2017enriching}. By mapping each sub-word to a $300$-dimensional vector, we can evaluate how the neural population encodes these units as continuous trajectories in a high-dimensional space.

\paragraph{Phoneme representations: One-Hot-Encoder.}
At the phonetic level, each of the $39$ English phonemes selected from the international phonetic alphabet (IPA) is represented using a one-hot encoding scheme.

\subsection{Modeling}
We employ two complementary linear modeling approaches to map the relationship between neural activity and speech features.

\paragraph{Encoding.}
Encoding is performed using the TRF to identify "speech-responsive" electrodes. This approach allows us to quantify the sensitivity of specific neurons to phonetic, syllabic or lexical variables across different time lags.
Given a neural signal $X \in \mathbb{R}^{T \times C}$ across $T$ time samples and $C$ channels, and a set of $F$ linguistic features $Y \in \mathbb{R}^{T \times F}$, we model the activity of each electrode as a linear combination of the features presented at various time lags $\tau$.
For each time step $t$, the predicted neural activity $\hat{X}_t \in \mathbb{R}^C$ is given by: $\hat{X}_t = W \hspace{0.1cm} \text{Concat}(Y_{t - \tau + 1} ... Y_t)$ where $W \in \mathbb{R}^{C \times (\tau \cdot F)}$ represents the learned weights (the TRF).

\paragraph{Decoding.}
Decoding analyses are based on ridge regressions to determine the information content available in the neural signal. We train decoders time-locked to the onset of phonemes, syllables and words. For each patient, analyses are performed both on individual Utah arrays and across all $4$ arrays.
We aim to predict a set of target linguistic features $Y \in \mathbb{R}^F$ from the population vector $X_t$ at a specific time $t$ relative to the event onset. 
The predicted features $\hat{Y}$ at time $t$ is modeled as: $\hat{Y} = W_t X_t$ where $W_t \in \mathbb{R}^{F \times N}$ are the learned weights and $N$ is the number of channels.
We implement ridge linear regression using scikit-learn \citep{pedregosa2011scikit}, with an alpha regularization per target dimension. Language representation features are standardized (zero-meaned and scaled to unit variance) per dimension and neural signals are robustly scaled (median-reduced and scaled according to the inter-quartile range) prior to model fitting.

\paragraph{Temporal generalization.}
To assess the stability and evolution of these representations, we extend the decoding analysis to a temporal generalization \citep{king2014characterizing} framework, where a decoder trained at time $t$ is tested on the neural activity at time $t'$. This allows us to construct a matrix of $t \times t'$ performance, where a diagonal pattern indicates a dynamic code where the representation changes over time, while a square pattern indicates a static code.
The prediction for a test time $t'$ using a decoder trained at time $t$ is formulated as: $\hat{Y}_{t \rightarrow t'} = W_t X_{t'}$
Temporal generalization can also be used to assess how well decoders generalize across successive events. To do this, we test whether each time-specific decoder (trained on $Y_t^{(i)}$ at time $t$) could accurately decode preceding or subsequent events ($Y_t^{(i-1)}$ or $Y_t^{(i)}$) in the test set.

\paragraph{Train-test procedure.} 
Each decoder is trained and tested within subjects using $10$-fold group cross-validation. Features are grouped by unique sentences in a group $k$-fold cross-validation, with $90 \%$ of the groups used for training and $10 \%$ for testing to assess decoding performance.

\paragraph{Metric.} 
Decoding performance is evaluated using Pearson's correlation coefficient ($R$) between true and decoded features, averaged across dimensions ($d$). Since language representations vary in space and dimensionality, we focus on comparing the relative changes in decoding performance.

\paragraph{Neuron tuning.} 
For the identification of neurons tuned to specific phonemes, syllables, or words, \Cref{fig2}C–E displays the specific item to which each neuron is tuned, along with other example events. However, the statistical tests were conducted using all available events.

\paragraph{Half-peak.} 
To determine if a representation emerges with the onset of an event, we calculated the "half-peak" of decoding scores. This is defined as the point at which decoding performance reaches $50 \%$ of its peak value compared to $5$ s after event onset.

\paragraph{Neural representation velocity.}
To quantify the evolution of the speech hierarchy over time, we compute the velocity of neural representations for phonemes, syllables and words. We extract the patterns associated with the decodability of each linguistic unit across time from the temporal generalization analyses \citep{gwilliams2022neural}. We then perform principal component analysis (PCA) and the velocity is assigned to the angles of the first principal component, which captures the primary axis of state-space evolution. This metric allows us to compare the relative speed of the neural evolution across different hierarchical levels.

\paragraph{Statistics.} 
To evaluate encoding and decoding scores, we employ a combination of non-parametric Wilcoxon signed-rank tests and Mann-Whitney $U$-test, and parametric $t$-tests and analysis of variance (ANOVA). ANOVA and $t$-tests are specifically applied when comparing mean decoding performance across cross-validation folds. To test the slope of the least-squares regressions, we use a Wald test with $t$-distribution. In the case of multiple comparisons across time points, electrodes or linguistic features, $p$-values are adjusted using the FDR correction as implemented in the MNE library \citep{gramfort2013meg}. All statistical analyses are performed using the Scipy library \citep{virtanen2020scipy}.

\section{Acknowledgements}
The authors would like to thank the BrainGate study team for their work and for open-sourcing the datasets.
Parts of this research were carried within the PhD funding AMX and the European Union's Horizon 2020 research and innovation programme under the Marie Skłodowska-Curie grant agreement No 945304 - Cofund AI4theSciences hosted by PSL University.



\clearpage
\newpage
\bibliographystyle{assets/plainnat}
\bibliography{main}

\newpage
\section{Supplementary Information} \label{Supplementary}
\renewcommand{\thefigure}{S\arabic{figure}}
\setcounter{figure}{0}

\begin{figure}[h]
    \centering
    \includegraphics[scale=0.4]{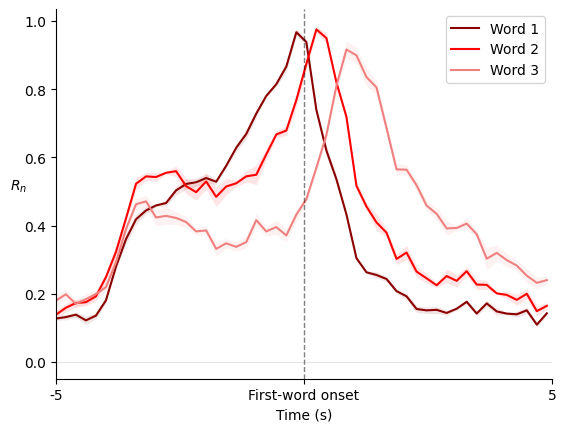}
    \caption{\textbf{Normalized decoding scores for Utah array 6v inf time-locked to sentence onset.} The decoding scores for the first three words of each sentence peak successively, corresponding to the temporal unfolding of the sentence. Notably, the first peak of these bimodal curves remains fixed in time across words, suggesting that it represents the neural response to the visual cue onset.}
    \label{supp1}
\end{figure}

\begin{figure}[h]
    \centering
    \includegraphics[scale=0.4]{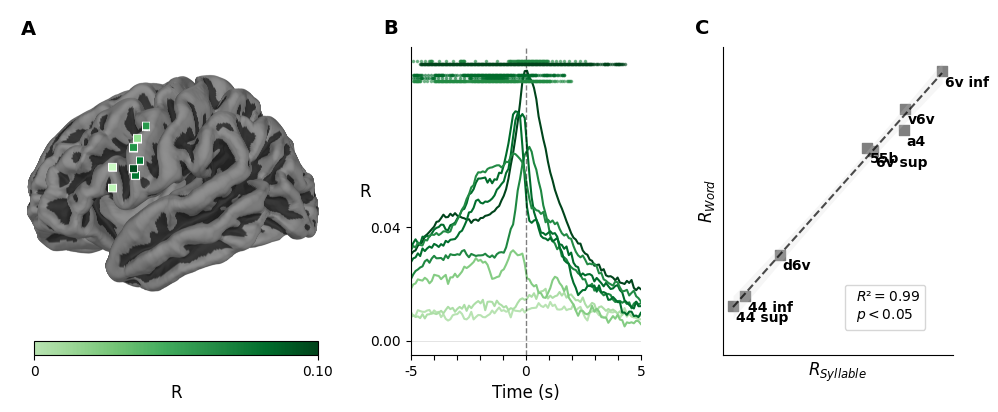}
    \caption{\textbf{High-resolution encoding and decoding of all syllables in the motor cortex and IFG.} \textbf{A.} Cortical map of all syllables peak decoding score within a $10$ s temporal window around onset, showing the eight Utah arrays.
    \textbf{B} Time-resolved decoding scores ($R$) for all syllables for all eight Utah arrays; significance is indicated by dots at each timestamp and array, based on one-sided $t$-tests across splits for $p < 10^{-11}$. Color coding is consistent with panel \textbf{B}.
    \textbf{C.} Linear regression of word versus all syllable peak decoding performance across arrays ($p < 5 \times 10^{-8}$, two-sided Wald test with $t$-distribution; $R^2 = 0.99$).}
    \label{supp2}
\end{figure}

\clearpage
\Cref{supp2} replicates the analyses from \Cref{fig2}, but includes syllables from all words, including monosyllabic ones. In \Cref{fig2}, phonemes and syllables show comparable decoding performance when only syllables from polysyllabic words are considered. However, when all syllables are included, syllable decoding performance aligns more closely with that of whole words. This overlap complicates the interpretation of syllabic feature analyses. In future work, it may be of interest to design sentences containing more polysyllabic words to disentangle this effect.

\begin{figure}[h]
    \centering
    \includegraphics[scale=0.3]{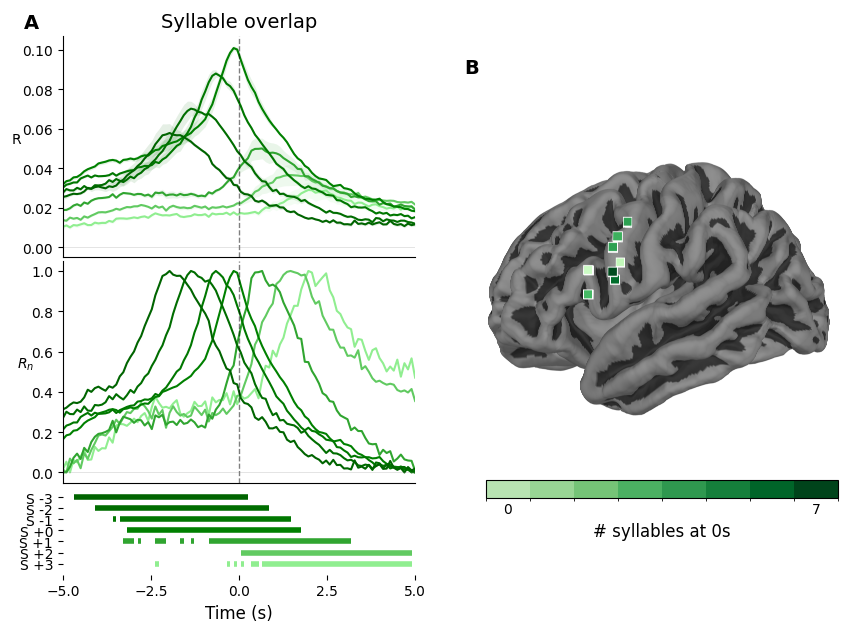}
    \caption{\textbf{Syllabic overlapping neural representations.} \textbf{A} Overlap of decoding scores in neural representations for all successive syllables aligned to the onset of the current event. These plots represent scores (top) and normalized scores (middle) for decoders trained on the $i^{th}$ ($i \in \{-3, -2, ..., 3\}$) syllable but time-locked on the $0^{th}$. Colored bars (bottom) represent the values above the median. Results are derived from all electrode data and averaged across patients, with error representing the SEM.
    \textbf{B.} Cortical maps showing the number of syllables (all) significantly decoded simultaneously at $t = 0$ s for each Utah array.}
    \label{supp3}
\end{figure}

\begin{figure}[h]
    \centering
    \includegraphics[width=\linewidth]{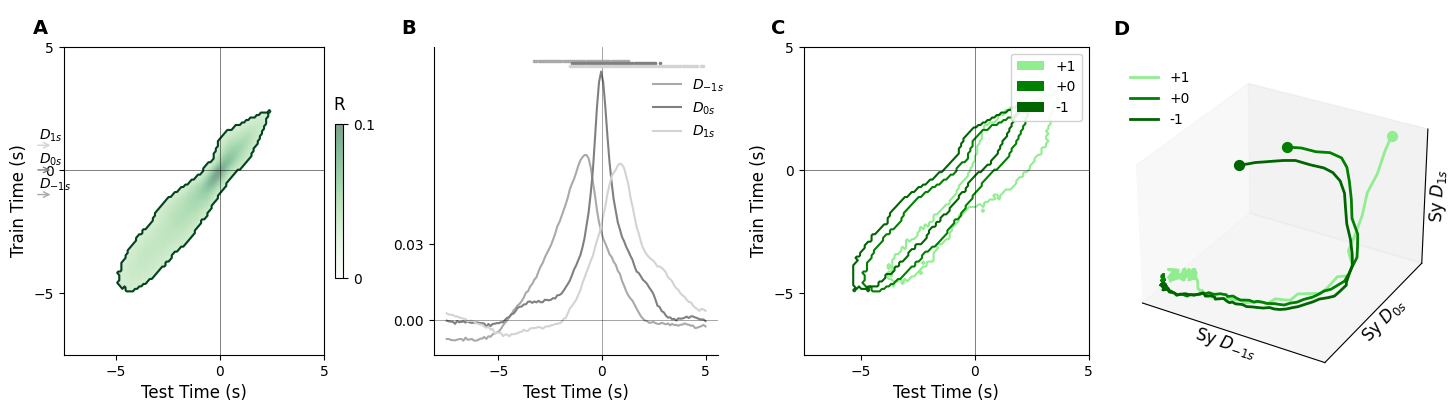}
    \caption{\textbf{Dynamic neural trajectories coordinate syllables.} \textbf{A.} Temporal generalization matrices for all syllables. Contours enclose scores exceeding the mean by $1.5$ standard deviations.
    \textbf{B.} Performance of decoders trained at specific time points ($D_{-1s}$, $D_{0s}$, $D_{1s}$) and tested across time for all syllables; significance is determined by two-sided Wilcoxon signed-rank tests across splits for $p < 10^{-10}$.
    \textbf{C.} Temporal generalization matrices for all syllables for successive speech units ($+1$, $0$, $-1$). Contours enclose scores exceeding the mean by $1.5$ standard deviations.
    \textbf{D.} $3D$ visualization of the neural code trajectories until $1.5$ s after all syllable onset for successive speech units ($+1$, $0$, $-1$). The axes represent normalized decoding scores for decoders trained at specific time points ($D_{-1s}$, $D_{0s}$, $D_{1s}$).}
    \label{supp4}
\end{figure}

\begin{figure}[h]
    \centering
    \includegraphics[scale=0.4]{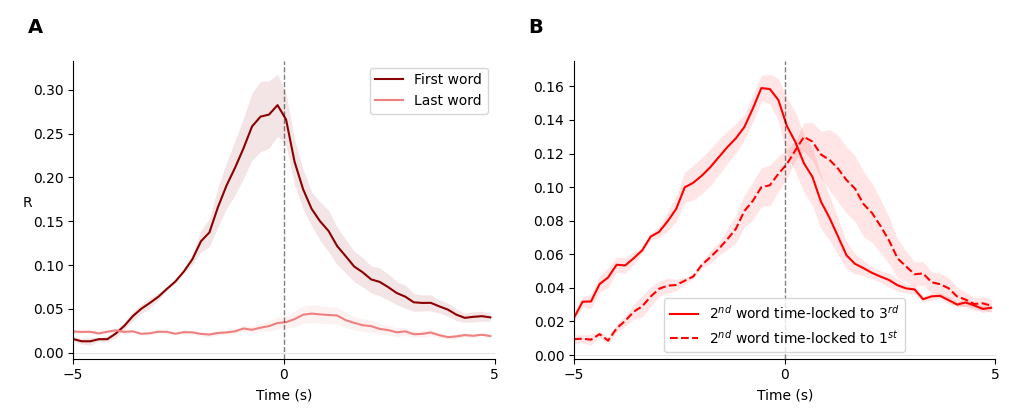}
    \caption{\textbf{Decoding performance in imbalanced dataset.} \textbf{A.} Decoding scores for neural representations of the first and last words in each trial sentence, time-locked to the onset of the first and last word, respectively. The results indicate that first words are highly decodable, while last words are poorly decodable. However, when decoding the next word, the first words must be excluded from analysis due to their position at the sentence boundary. Conversely, when decoding the previous word, the last words must be excluded. This exclusion process accounts for the higher decoding scores observed for previous words compared to next words.
    \textbf{B.} An example focusing on the second word, analyzed with time-locking to both the previous and next words. The decoding scores exhibit greater similarity and the remaining difference likely stem from our use of word-onset time-locking. This approach ensures superior alignment for preceding words relative to subsequent ones.}
    \label{supp5}
\end{figure}

\clearpage
In \Cref{supp3} and \Cref{supp4}, the analyses from \Cref{fig3} and \Cref{fig4} are replicated with all syllables, again demonstrating strong similarities between the all-syllable and word analyses.

\clearpage

\end{document}